\documentstyle[12pt,epsf]{article}
\begin{document}
\begin{center}
{\ {\Large {\bf Transversal Flow of Pions as a Consequence}}}\\
\vskip 0.1cm
\vskip 0.2cm
{\ {\Large {\bf of Rescattering Process }}}\\
\vskip 0.4cm
{\ {\small {\bf (in non-central Pb+Pb 158GeV/n collisions)}}}\\

\vskip 0.5cm

{\sl Peter Filip\footnote{Permanent address: Institute of Physics, 
SAS, Bratislava, SK-842 28}}\\
\vskip 0.25cm
Faculty of Mathematics and Physics, Comenius University \\
842 15 Bratislava, Slovak Republic \\
E-mail: filip@savba.sk \\

\vspace{0.32cm}
\end{center}

\begin{abstract}
Aim of this work was to test the idea of J.-Y.Ollitrault
about the parallel squeeze-out type of transversal flow in mid-rapidity
region \cite{Det}. For this purpose we have performed a computer simulation
of the expanding pion gas created in non-central Pb-Pb 158GeV/n
collisions. A squeeze-out type of asymmetry parallel to impact parameter
in azimuthal distribution of pions is found and studied. The asymmetry
is explained as a consequence of geometry of non-central collisions and
the rescattering process.
\newline 

\hskip-0.6cm PACS numbers: 25.75
\end{abstract}
\vspace{0.2cm}

\begin{center}
{\bf 1. Introduction}
\end{center}
Azimuthal asymmetries in transverse momentum distributions of particles
measured in relativistic heavy ion collisions (HIC) were observed
for the first time by plastic-ball detector in Berkeley \cite{Bev}.
Since that pioneering experiment the asymmetries in relativistic HIC were
measured also at higher energies \cite{Barrette,AGS}.
Typical types of asymmetries (bounce-off and squeeze-out) are explained 
as a consequence of collective behaviour of nuclear matter.
This understanding of the origin of azimuthal asymmetries in non-central
HIC succesfully explains most of the experimental data. 

However at AGS experiment with Au+Au 11.4 GeV/n collisions E877 collaboration
reported \cite{Barrette} about the unclear origin of a new type squeeze-out
effect parallel to impact parameter. It seems that E877 collaboration has
found the first experimental indications of the effect predicted by 
J.-Y.Ollitrault \cite{Det}. This type of squeeze-out effect results
from the interaction of produced particles among themselves.

In this work we investigate asymmetry in transverse momentum distribution
of pions created in the process of HIC. No collective
behaviour of nuclear matter is supposed in the calculation. Asymmetry
results from the geometry of non-central HIC and the rescattering process.

Paper is organized as follows: In Section 2 we give an intuitive
explanation of the origin of the asymmetry. Description of the computer
simulation is given in Section 3. In Section 4 we present results of
our simulation - dependence the asymmetry on impact parameter and also
other features of the studied effect.
We finish this work with short summary and conclusions.

\begin{center}
{\bf 2. Asymmetry as a consequence of rescattering}
\end{center}

Usual explanation of the azimuthal asymmetries in momentum distributions of
particles in HIC is based on the collective behaviour of nuclear matter.
Hydrodynamical models \cite{BJ,Ani} describe the collision process of
HIC using the equation of state for nuclear matter.
Substantial assumption of these models is a thermalisation process.
As we shall see the existence of thermal equilibrium is not required
for the existence of our type of asymmetry.

Number of secondary particles created in HIC at present SPS energies 
(158GeV/n) is substantially larger than the number of nucleons contained 
in the colliding nuclei.
Therefore the situation is different than that at lower energies and the
interaction of the secondaries among themselves becomes significant.
As a signature of the rescattering process among the produced pions we
regard the $\vec p_t$ dependence of transverse size parameters $R_t$ 
\cite{NA44}
extracted by HBT technique \cite{Zajc}. 
The decrease of $R_t$ with increasing $p_t$
was explained as a consequence of the rescattering of pions in the simulation
\cite{Hum} where the central S-Pb 200 GeV/n collisions were studied.

We think that the rescattering of secondaries can demonstrate itself in the
experimental data also as a transverse flow of pions in non-central 
collisions.
For the existence of this effect no collective behaviour of nuclear matter
is necessary. Therefore the phenomenon can be observed also in the future
HIC experiments (RHIC, LHC) where nuclear transparency region is expected
to be reached.

In this section we explain the origin of the predicted transverse
flow of pions as a consequence of the geometry of non-central HIC and the 
rescattering process. On Fig.1a we show the geometry of non-central A+A 
collision
in transverse plane. Secondary particles i.e. also the pions are created
mainly in the overlapping region where nucleon-nucleon collisions happen.
Shape of the overlapping region depends on impact parameter $b$ and is
azimuthally asymmetric in transverse plane.
Since we do not consider any collective behaviour of nuclear matter it is
natural to assume that the initial distribution of pions in transverse momenta
$\Psi (\vec p_t)$ is azimuthally symmetrical:
$
\Psi ^S(\vec p_t)=\Psi (|\vec p_t|)
$.
Because of the simplicity of explanation we do not write longitudinal
components of $\vec p$ and $\vec x$ in this section however the simulation
described in the next section is performed in 3 dimensions.

Denoting the space distribution of the points of creation of pions
by $\Phi (\vec x_t)$ we have the following initial condition for the pion
gas created in HIC:

$$
X (\vec x_t,\vec p_t)=\Phi ^A(\vec x_t)\cdot \Psi ^S(\vec p_t)
\eqno{(1)}
$$

Because of the rescattering process the original $\vec x-\vec p$ 
non-correlated distribution (1) becomes $\vec x-\vec p$ correlated \cite{Acta}
and (as we shall see)
the asymmetry in $\Phi ^A(\vec x_t)$ {\it leaks} into asymmetry in the
resulting transverse momentum  distribution.
$$
\Psi ^{S}(\vec p_t)\cdot \Phi ^A (\vec x_t) \rightarrow
X ^{A}(\vec p_t,\vec x)
\eqno{(2)}
$$
where $ ^A$
denotes asymmetry in distribution $X (\vec p_t,\vec x)$ in momentum.

This effect can be understood in the following way:
Let us have two groups of pions:
A group of parallel pions with the momentum parallel to $\vec b$
and the group of orthogonal pions with the momentum orthogonal to $\vec b$.
Because of the asymmetry in $\Phi ^A(\vec x_t)$ the probability of collision
of the orthogonal pion is (in average) higher than the probability of 
collision for paralell pion (see Fig.1b). 

\vskip0.4cm
\centerline{\epsfxsize=12.5cm\epsffile{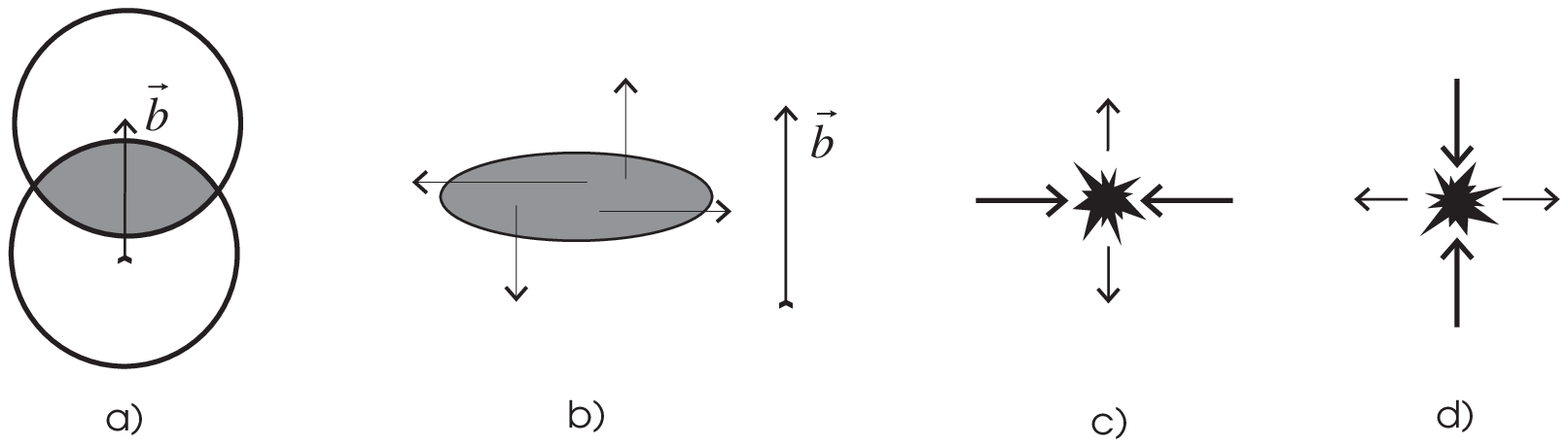}}
\vskip1.7pt
\centerline{\small {\bf Fig.1} Geometry of non-central collision in
transversal plane}
\vskip0.2cm

The reason for this is very simple:
Number of pions located in the direction orthogonal to $\vec b$
is higher because of the bigger size of initial "cloud" of pions  in
this direction. Thus during the rescattering process orthogonal pions
collide more likely than the parallel pions and consequently collisions
of type Fig.1c are more frequent than the collisions of type Fig.1d.
Such conditions lead to the excess
in the number of paralell pions.

This is undoubtedly a non-equilibrium process. Resulting asymmetry in the
momentum distribution of pions freezes in the considered pion gas because
of the expansion of the system.

Azimuthal asymmetry in transverse momentum distribution of pions after
the rescattering process can be measured by detectors as the excess of
pions in the direction paralell to the impact parameter. As we shall
see in the next sections predicted asymmetry should  demonstrate itself
as a second order asymmetry in the fourier analysis of transverse flow 
\cite{Zhan}.

On Fig. 2 we show the result of a toy simulation  of the effect described.
Initial distribution of pions in momenta was taken from the central
S-Pb 200 GeV/n HIC simulation \cite{Acta,Hum} and therefore it was azimuthally
symmetric. Initial distribution of pions in $\vec x_t$ space is artificially
asymmetric as it is shown on Fig. 2b. After 20fm/c of the time evolution
the rescattering process leads to the asymmetry in transverse momentum
distribution as it is clearly seen from Fig. 2c.

\vskip0.2cm
\centerline{\epsfxsize=8.8cm\epsffile{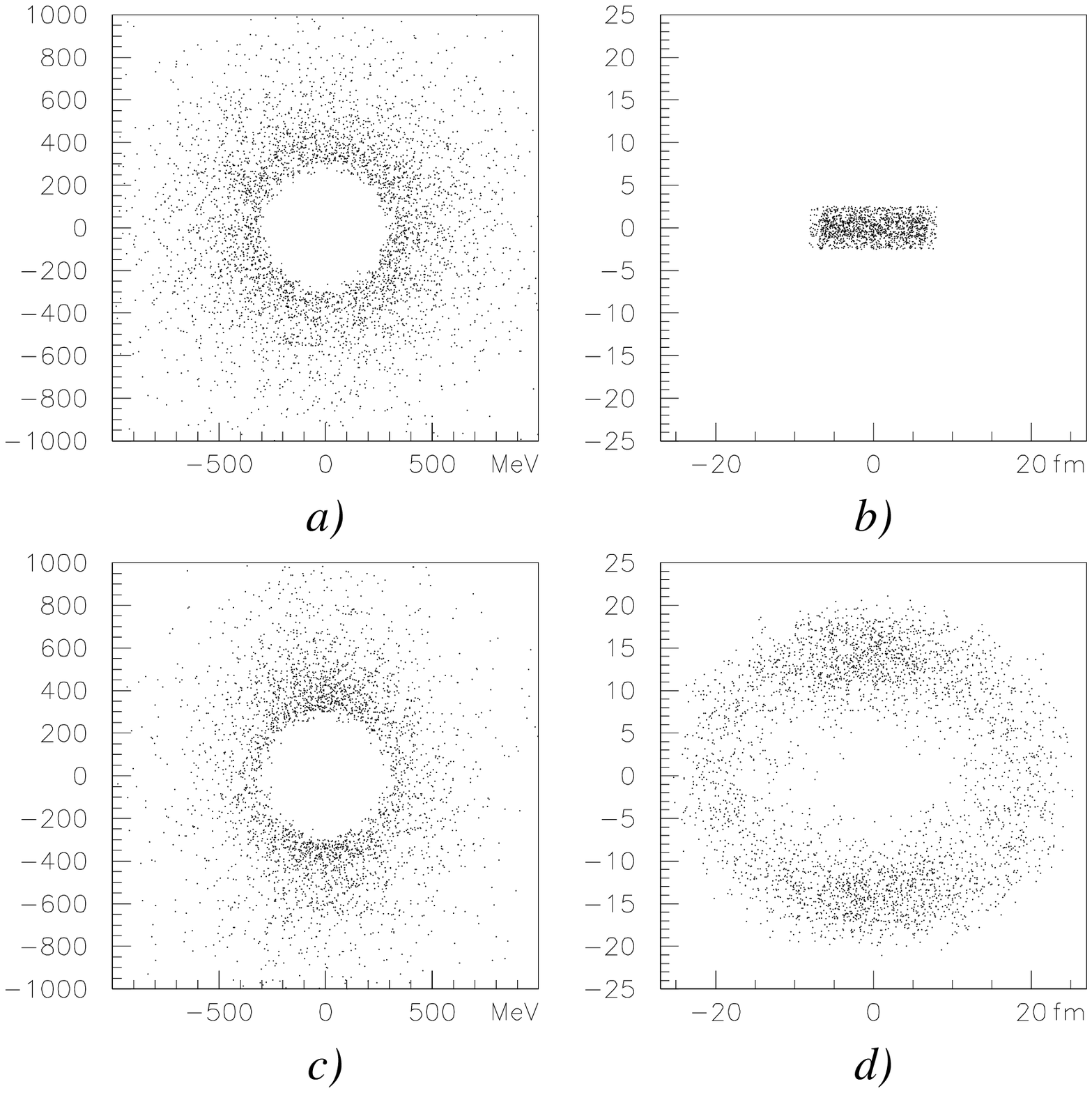}}
\vskip1.7pt
\centerline{\parbox{12cm} {\small {\bf Fig.2} Toy simulation of the effect.
{\it a)} is $\vec p_t$ distribution before the rescattering. {\it c)} is
$p_t$ distribution after 20fm/c of rescattering process. Cut in transverse
momentum $p_t > 300MeV$ is applied to enhance the visual effect of the
asymmetry. {\it b)} is $\vec x_t$ distribution before the rescattering process
and {\it d)} after the 20 fm/c of expansion.
}}
\vskip0.2cm

Simulation of the expanding pion gas for "real" Pb-Pb 158 GeV/n non-central
collisions is described in the next section.

\begin{center}
{\bf 3. The Simulation}
\end{center}

For the simulation of the expanding pion gas created by "real" Pb-Pb 158 GeV/n
non-central collisions two independent programs were used:

1) Cascading generator \cite{PZ} which generates the 
initial momenta and positions of pions as a result of independent 
nucleon-nucleon collisions.

2) Rescattering program \cite{Acta} which simulates time evolution of the
interacting of pions created in HIC.

First we shall describe main interface structure of the simulation shown on
Fig.3. Pb-Pb 158 GeV/n collisions were simulated by the cascading generator 
(CG) for
random orientations and selected values of impact parameter $\vec b$.
Information about initial momenta, time and place of the creation of pions
$\pi (\vec x,\vec p,t)$
produced by CG was used as input for the rescattering program \cite{Acta}.
Final momenta of the interacting pions were selected from the output of the
rescattering program and analyzed for transversal asymmetries. Orientation of
the impact parameter was not supplied by CG and its determination
is described in subsection 3.3.

\vskip0.4cm
\centerline{\epsfxsize=13.7cm\epsffile{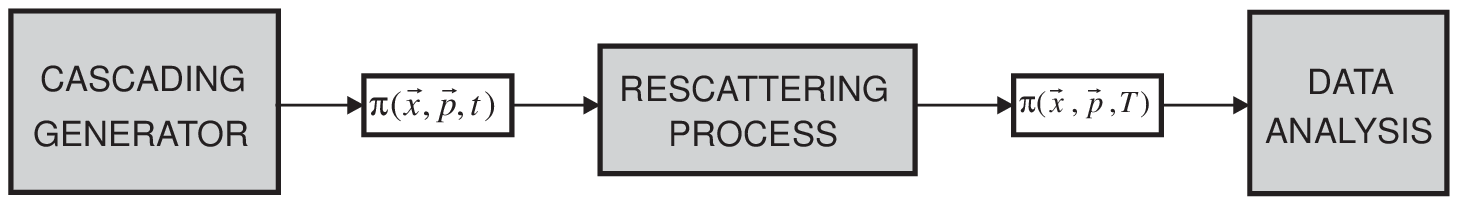}}
\vskip1.7pt
\centerline{\small {\bf Fig.3} Main interface structure of the simulation}
\vskip0.2cm

\vskip0.9cm
{\bf 3.2 Rescattering process }
\vskip0.4cm

Description of internal structure of the program used for the simulation
of rescattering process can be found in \cite{Acta}. 
This program was build according
to the description of the simulation of central S-Pb 200 GeV/n collisions 
\cite{Hum}.
Results obtained by program \cite{Acta} for the central S-Pb collisions are
in agreement with the results obtained in \cite{Hum}. We shall sketch here 
just main features of our program.

Pions are treated as point-like objects in the simulation,
position and momentum of each pion is known during the simulation. Pions
move in small time steps $\Delta t=0.1fm$ as a free particles. Collisions
happen if two pions appear to be at a distance smaller than the critical 
distance
$D_k$ which is determined from the isospin averaged total elastic 
cross-section

$$
D_k=\sqrt {\sigma(s)/\pi}
\eqno{(3)}
$$

Momenta of pions after the collision are determined in CMS
of the pair according to the isospin averaged differential cross section
$$
d\sigma /d\Omega = a(s) + b(s) \cos ^2(\theta)
\eqno{(4)}
$$
(Numerical values of the functions $a(s),b(s)$ are calculated from data
\cite{Prak,Esta}.)

Then new momenta are transformed back to the global frame of the simulation.
Test for the relative distance is performed for every pair of pions in each
time step. Not all of pions evolve in time at the beginning of simulation.
Cascading generator produces also the information about the time of creation 
for
each pion therefore pions are tested for the existence in the global frame of
simulation.

\vskip0.7cm
\centerline{\epsfxsize=7.7cm\epsffile{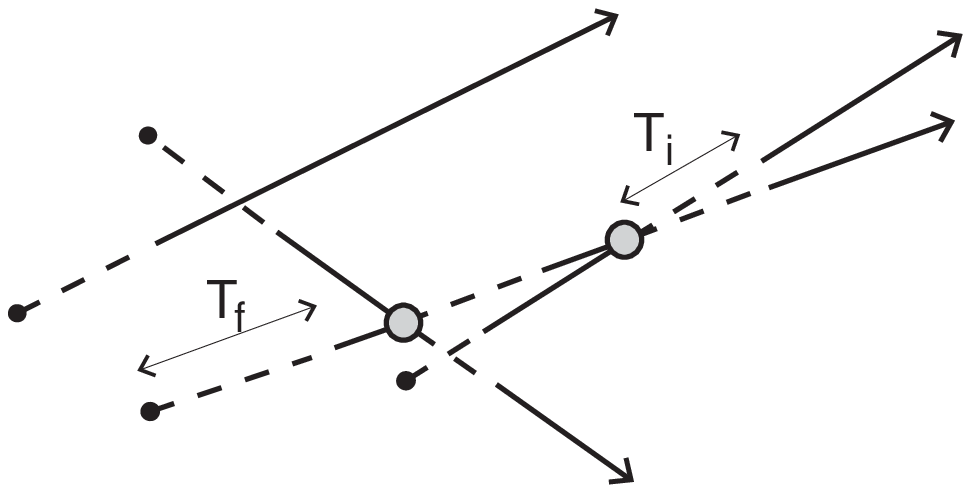}}
\vskip1.7pt
\centerline{\parbox{12cm} {\small {\bf Fig.4} Dynamics of $\pi \pi$ 
collisions.
Collisions can happen only if both pions are allowed to interact 
(solid lines).
}}
\vskip0.3cm

Moreover existing pions are restricted from the interaction for
time $T_f$ - formation time \cite{Pisut} after their creation and also
for time $T_i$ after each collision (see description in \cite{Acta}). These
two phenomenological parameters allow to influence total number
of collisions in the simulation and also some features of the expanding
of pion gas (see Tab.1).

\vskip0.9cm
{\bf 3.3 Data analysis}
\vskip0.4cm
Events of the exact values of impact parameter $b=3,5,7,9,11${\it fm}
were used for the analysis.
Approximate orientation of impact parameter was determined from output
of the cascading generator. Cascading generator \cite{PZ} simulates 
A-A collision with the
target nucleus at the center of coordinate system. Therefore the initial
positions of pions which lie mainly in the overlapping region of nuclei are
shifted from the center of coordinate system in the direction of impact
parameter (see Fig.5).

A procedure close to Danielewicz - Odyniec method \cite{Dan} was used 
for determination of the approximate orientation of impact parameter.
For each event vector
$$
\vec Q = \sum _n \vec x_t^i
\eqno{(5)}
$$
($\vec x^i_t$ are initial positions of created pions in transversal plane)
was constructed and its orientation was used as the orientation of impact
parameter $\vec b$. Final momenta of pions after the rescattering
process were rotated to have the same orientation of the approximate impact
paramenter.

\vskip0.2cm
\centerline{\epsfxsize=13.7cm\epsffile{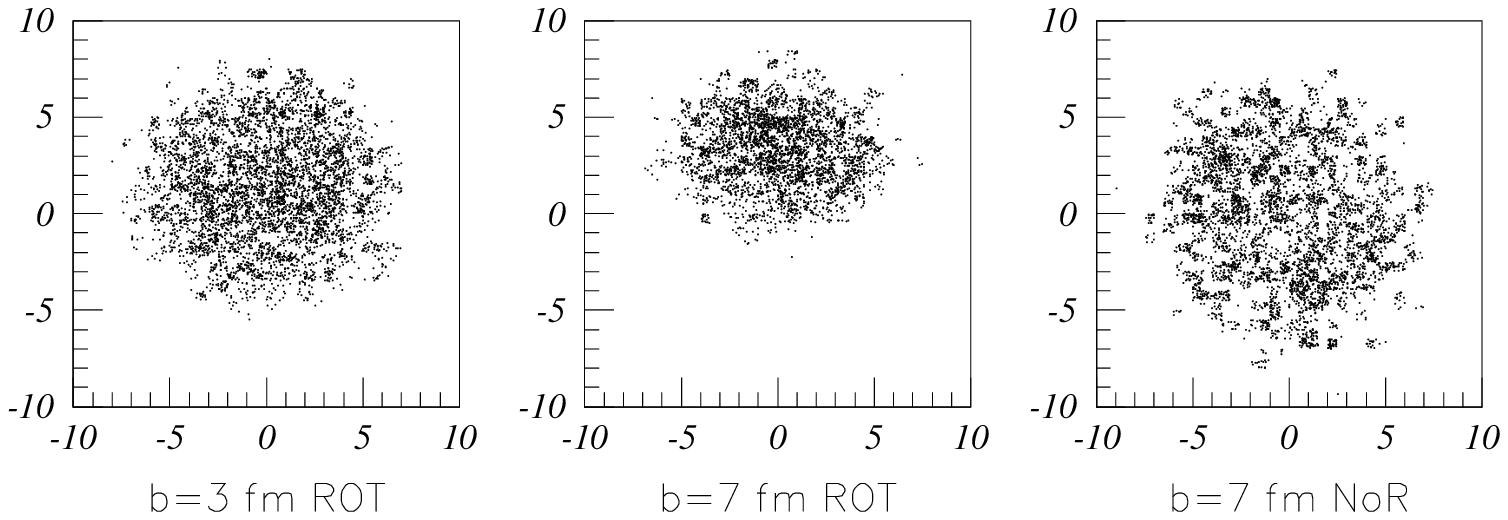}}
\vskip1.7pt
\centerline{\parbox{12cm} {\small {\bf Fig.5} Initial positions of pions 
in transversal plane obtained from the cascading generator. 10 events for
b=3fm and b=7fm are rotated to have the same orientation of $\vec b$. On
figure b=7fm NoR the rotation is not performed.
}}
\vskip0.2cm

Rotated momenta of pions were used for fourier type of analysis of transversal
flow \cite{Zhan}. For each pion azimuthal angle of momentum in respect to
the impact parameter was determined and added to histogram of azimuthal
distribution $R(\phi )$ (see Fig.6). 
Because of the known character of asymmetry and
the rotation of events into the direction with the same orientation of
impact parameter we have fitted the normalized histograms $R^N(\phi)$
to the function
$$
R^N(\phi )= 1 + S_2 \cos (2\phi )
\eqno{(6)}
$$
Results of the analysis are presented and discussed in the next sections.

\begin{center}
{\bf 5. Results}
\end{center}

First the presence of asymmetry was tested on the set of 24 
artificial S-Pb 200 GeV/n 
asymmetrical events.
A strong final asymmetry in azimuthal distribution of
pions in transversal momentum is visible directly from Fig.2. 
We have tested also dependence of the strength of the effect
on the dynamics of rescattering process. 
Normalized histograms of $R(\phi )$ distribution were fitted to 
function (6) for different values of 
parameters $T_f$ and $T_i$. 
The results obtained are summarized in Table 1.

\vskip0.2cm
\centerline{\epsfxsize=13.7cm\epsffile{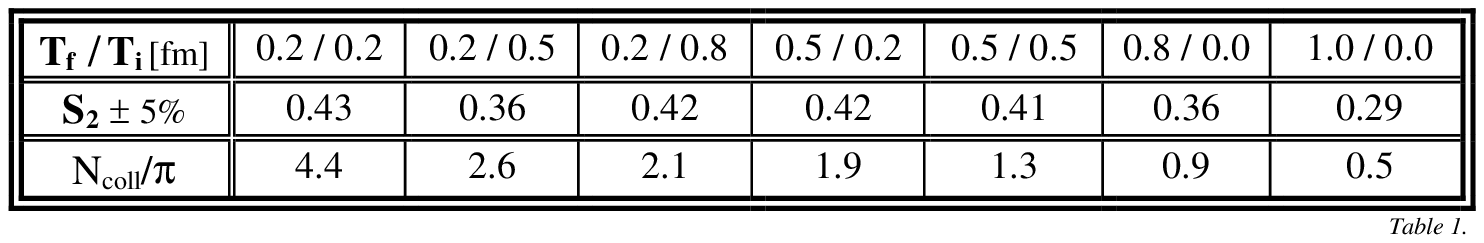}}
\vskip1.7pt
\centerline{\parbox{12cm} {\small {\bf Tab.1} Results of fit for 24 artificial
S-Pb events. N$_{Coll}$/$\pi$ is average number of collisions per $\pi$ during
the rescattering process. Total number of pions in one event was 700.
}}
\vskip0.2cm

Then "real" non-central Pb-Pb 158 GeV/n collisions were studied
using the output of cascading generator \cite{PZ}.
We have run the program for five sets of 10 events 
with impact parameter $b=3,5,7,9,11 fm$
in order to study dependence of the asymmetry on the impact parameter
and additional 20 events for $b=7 fm$ were run in order to verify
whether the statistics of 10 events per set was sufficient\footnote{For 
$b=11$fm 20 events were run because of low multiplicity of pions in
these events.}.
Pions in rapidity range $\langle -1,1 \rangle$ were
selected for the analysis. Events were rotated to have the same orientation
of impact parameter using the procedure described in section 3.3.
On Fig.6 we show the histograms of distribution $R(\phi )$
for the ROT - rotated (to the same orientation of $\vec b$) and non-rotated
(NoR) events for b=3fm and b=9fm.
Normalized distributions $R^N(\phi )$ were used for the
fit.

\vskip0.2cm
\centerline{\epsfxsize=13.7cm\epsffile{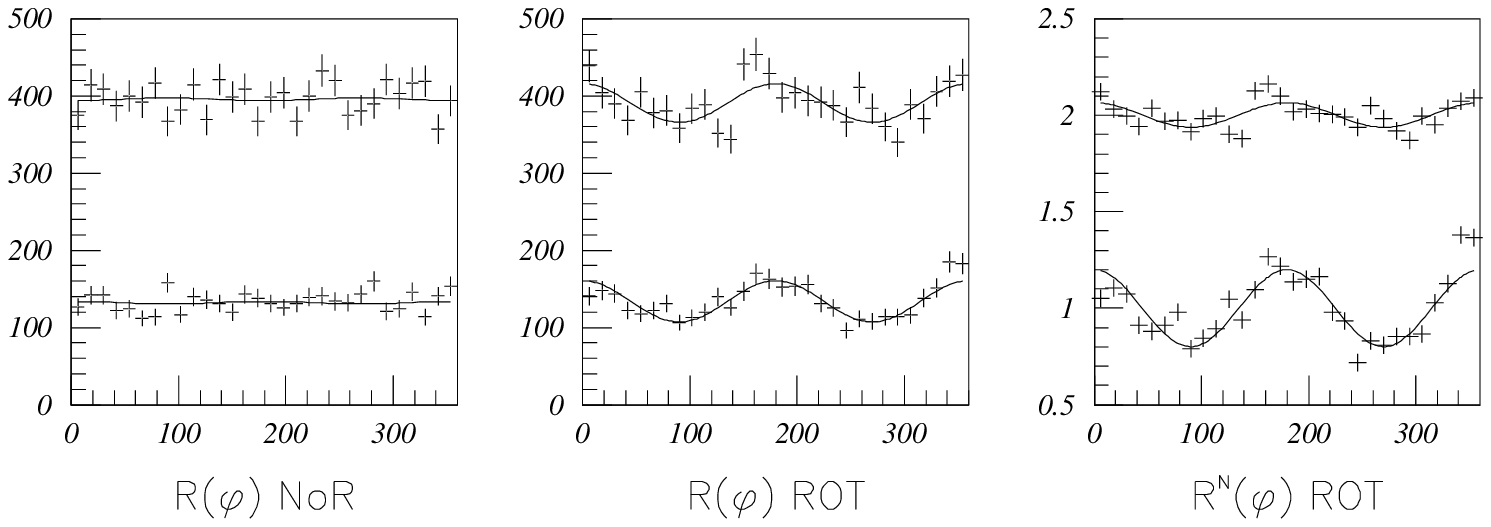}}
\vskip1.7pt
\centerline{\parbox{12cm} {\small {\bf Fig.6} 
Histograms of the azimuthal distributions of pions for b=3fm and b=9fm.
Numbers of particles in the bins of histogram for b=3fm are higher
than in the case b=9fm because of the higher total multiplicity of pions
in the collisions b=3fm. Normalized histogram $R^N(\phi )$ for b=3fm
is artificially shifted up.
}}
\vskip0.2cm

Results of the fit are summarized in Tab.2. Simulation of
20 additional events for b=7fm showed that the statistics
of 10 events is  sufficient for the qualitative analysis we have
performed. Main result of our simulation is visible directly
from Fig.7: Asymmetry increases with impact parameter in range 
$3-9${\it fm} and it falls down for $b>9${\it fm}.
Turnover point of $S_2$
lies between 7 and 11 {\it fm}. We think that the position of the turnover 
point depends on $T_f, T_i$ parameters used in the simulation.

Rapidity dependence of the asymmetry was analyzed for events with 
impact parameter 
b=7{\it fm}. Our small statistics allowed us to perform only a very rough
analysis of rapidity dependence of $S_2$ coefficient. Results in the
following table are averaged  for forward and backward rapidities arround
central rapidity what is in our simulation $Y_c=0.0$.

\begin{table}[h]
\begin{center}
\begin{tabular}{c|c|c|c}
$|{\mbox Y}|$ & 
$\langle 0,1 \rangle $ & $\langle 1,2 \rangle $ & $\langle 2,3 \rangle $\\
\hline 
$S_2$   &      0.14        &         0.15        &         0.9 \\
\end{tabular}
\end{center}
\end{table}

We have tried to find also first order asymmetry (bounce-off) in our data
which would demonstrate itself as a non-zero value of $S_1$ coefficient
in the fourier expansion:
$$
R(\phi)=A_0\cdot (1+S_1\cdot \cos (\phi )+ S_2\cdot \cos (2\phi ))
\eqno{(8)}
$$
For events $b=7${\it fm} we have obtained following results: 

\begin{table}[h]
\begin{center}
\begin{tabular}{c|c|c}
Y & $\langle -3,-1 \rangle $ & $\langle 1,3 \rangle $ \\
\hline
$S_1$ before & $+0.007 \pm 0.01$ & $-0.003 \pm 0.01$  \\
\hline
{$S_1$ after\ \hspace{3.6pt}}  & $+0.016 \pm 0.01$ & $-0.02 \pm 0.01$   \\
\hline
$\langle p_x \rangle $ in MeV & 2.22 & -1.96 \\
\hline
$\langle p_y \rangle $ in MeV & 0.06 & -0.06 \\
\end{tabular}
\end{center}
\end{table}

$S_1$ {\it before} is calculated from the output of cascading generator
and $S_1$ {\it after} was computed for events after the rescattering 
process (see Fig.3). 
Slight signature of the first order
asymmetry is present in the output of cascading generator 
what can be a consequence of the cascading of nucleons in spectator matter
included in the generator \cite{PZ}. However this effect is not
above the statistical errors in our set of evets b=7{\it fm}. 

Calculation of average $\langle p_x \rangle $ and $\langle p_y \rangle $ of 
pions in the resulting events b=7{\it fm} confirms that the first order 
asymmetry is very weak in our data.

We have computed $R_p$ parameter used in the study of squeeze-out effect at 
BEVALAC/SIS energies \cite{Gut}
$$
R_p={{\langle p_x^2\rangle - \langle p_x \rangle ^2 }
\over {\langle p_y^2\rangle - \langle p_y \rangle ^2}}
\eqno{(9)}
$$
where $x$ direction is parallel to impact parameter. For b=7{\it fm} 
our $R_p$ of rotated (ROT) and non-rotated (NoR) events is:
$$
R_p^{{ROT}}=1.39 \quad \quad R_p^{{NoR}}=0.97
\eqno{(10)}
$$

An interesting behaviour of asymmetry coefficient $S_2$ was found
in our data analysis. As it is shown in Tab.2 coefficient $S_2$
is significantly higher for high $p_t$ 
pions ($p_t > 300$MeV).

\vskip0.5cm
\centerline{\epsfxsize=13.8cm\epsffile{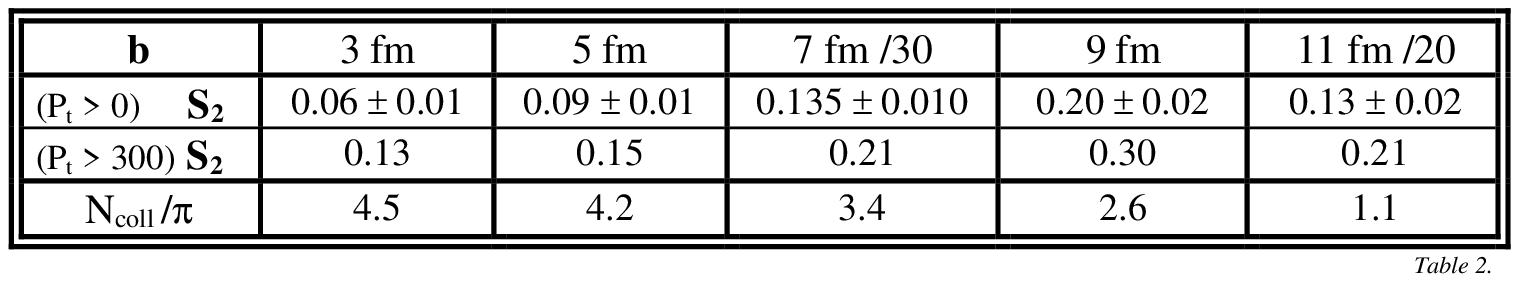}}
\centerline{\parbox{12cm} {\small {\bf Tab.2} Asymmetry coefficients for
Pb-Pb 158 GeV/n non-central events. 7 fm/30 means that the fit was performed
for 30 events.
}}
\vskip0.4cm

We have tested whether this behaviour is present also in $E_t$ sensitive
analysis. For this purpose we have filled histograms of $R(\phi)$ distribution
by the value $E_t$ for each pion. $P_t$ dependence of $S_2$ coefficient
was still present in the results.

\begin{center}
{\bf 5. Summary and Conclusions}
\end{center}

We have studied asymmetry in the azimuthal transverse momentum distribution
of pions for non-central Pb-Pb 158 GeV/n collisions. For the presence of this
type of asymmetry no collective behaviour of nuclear matter is
necessary. It is a consequence of the geometry of non-central collisions
and the rescattering among the pions.

The simulation showed
that the asymmetry increases with the impact parameter in the
range up to $9fm$. Turnover point is located between $7-11fm$ in our data.
Decrease of asymmetry coefficient $S_2$ for $b=11fm$ is most likely 
a consequence of low number of pions participating in the rescattering
process combined with small total size of
the overlapping region in comparison to formation path $L_f \simeq c\cdot T_f$
of pions. Non-interacting pions in the ''formation time stage`` do not feel 
the asymmetric shape of the overlapping region of colliding ions and therefore
the asymmetry decreases. If this is the main reason of the decrease of 
asymmetry at peripheral collisions then the position of the turnover point 
can be sensitive to the value of formation time parameter $T_f$ used in 
the simulation.

We think the effect studied in this work was already confirmed experimentally
by NA49 collaboration \cite{Snow} using Ring Calorimeter setup \cite{Na49}.

Unexpected $p_t$ dependence of the  $S_2$ asymmetry coefficient was
found in the analysis of our data. This seems to be an interesting prediction
however the origin of this $p_t$ dependence is not clear at present.
It can be a consequence of our scenario of rescattering process and therefore
it does not need to appear in experimental data. Results obtained from 
NA49 TPC could  answer this question. We hope some data will be presented 
during QM'96 conference.

\vskip0.2cm
\centerline{\epsfxsize=10cm\epsffile{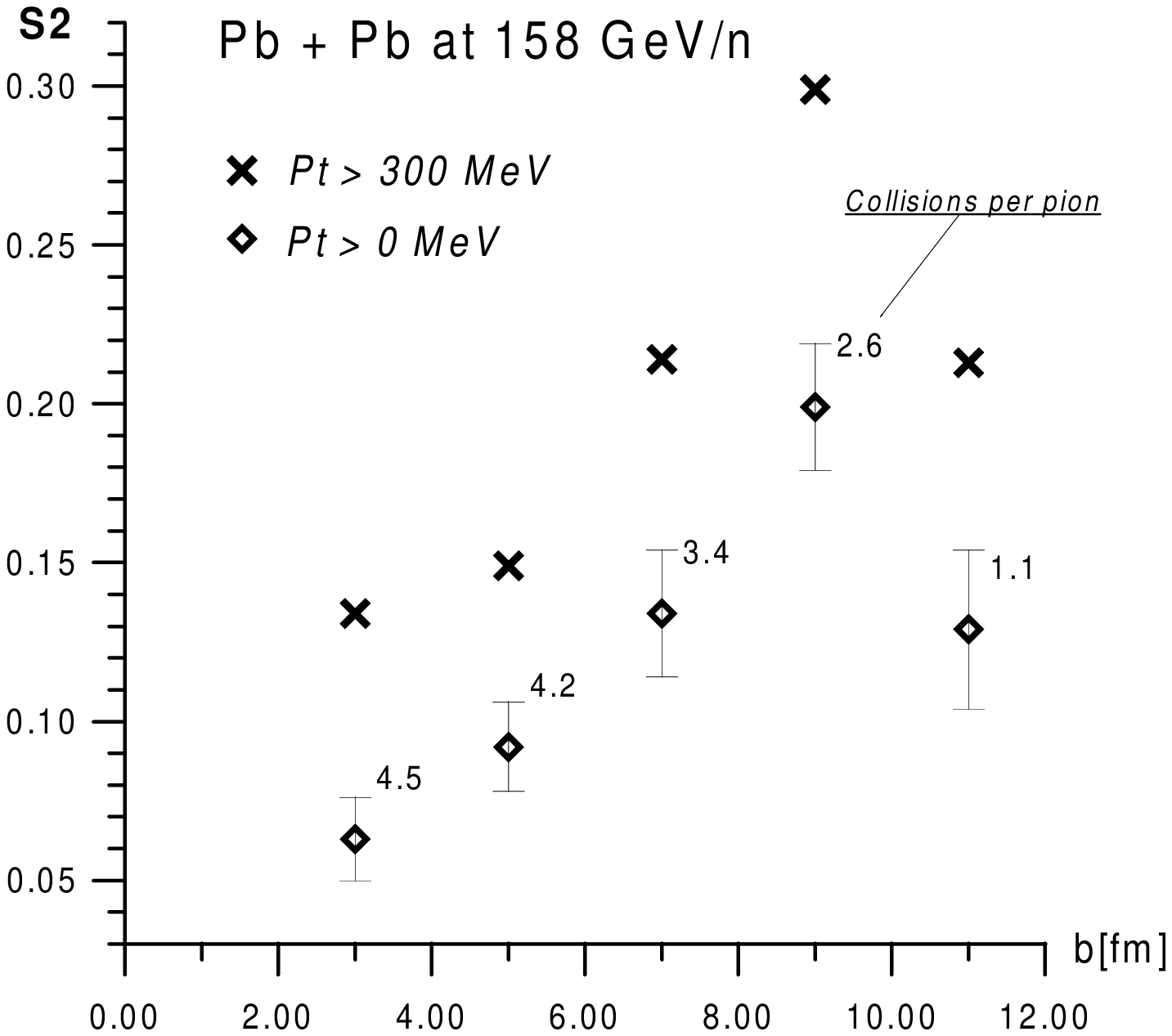}}
\centerline{\parbox{12cm} {\small {\bf Fig.7}  Dependence of $S_2$ coefficient
on impact parameter. For high-$p_t$ pions the $S_2$ coefficient is 
systematically higher than in the case of full $p_t$ range pions.
}}
\vskip0.2cm

At the future HIC experiments the multiplicities of secondaries will
be much higher. Because of the nuclear transparency phenomenon
the mechanism responsible for the transversal flow of nucleons at 
lower energies (up to 10GeV/n)
can play a little role. In this case the studied effect can be 
substantial for the transversal flow phenomenon
at HIC on  RHIC and LHC.

\begin{center}
{\bf Acknowledgements}
\end{center}

Author is indebted to prof. J.Pi\v s\'ut for supervising
this work. Special thanks are directed to P.Z\'avada - the author
of the Cascading Generator for the possibility to use output of
his program. We express our gratitude
to member of NA49 collaboration T.Wienold 
for the important comments and suggestions.

This work was originally motivated by the discussion with 
J.-Y.Ollitrault. We thank to J.-Y.Ollitrault also for the subsequent
fruitful discussions.

\newpage

\end{document}